\documentclass[runningheads]{llncs}
\usepackage[T1]{fontenc}

\usepackage{cite}
\usepackage{hyperref}
\usepackage{url}
\usepackage{subfig}
\usepackage{amsmath,amssymb,amsfonts}
\usepackage{graphicx}
\usepackage{textcomp}
\usepackage{xcolor}
\usepackage{multirow}
\usepackage{bbding}
\usepackage[linesnumbered,ruled,vlined]{algorithm2e}
\usepackage{algpseudocode}
\usepackage{booktabs}
\usepackage{array}
\usepackage{orcidlink}
\let\oldnl\nl
\newcommand{\nonl}{\renewcommand{\nl}{\let\nl\oldnl}}

\begin{document}

\title{Adaptive Cyber-Defense for Evolving Attacker Capabilities using Reinforcement Learning and Behavior Trees}
\titlerunning{ACD for Evolving Attacker Capabilities using RL and BTs}
\author{
    Nicholas Potteiger \orcidlink{0009-0005-0406-0355} \Envelope
    \and
    Ankita Samaddar\orcidlink{0000-0001-6154-7673}
    \and
    Xenofon Koutsoukos\orcidlink{0000-0002-0923-6293}
}
\authorrunning{Potteiger et al.}
\institute{
    Vanderbilt University, Nashville, TN, USA\\
    \email{
        nicholas.potteiger@vanderbilt.edu,
        ankita.samaddar@vanderbilt.edu,
        xenofon.koutsoukos@vanderbilt.edu
    }
}

\maketitle

\begin{abstract}
Autonomous cyber-defense agents trained using reinforcement learning (RL) are increasingly adopted for defending enterprise computer networks. However, existing approaches may become brittle as attacker capabilities evolve and enable new attack strategies. In this work, we study the problem of adapting defender policies learned in CAGE Challenge Scenario 2 (CC2), a benchmark enterprise network defense scenario, to an extended network scenario with new attacker capabilities represented by new attacker actions. Effective attack strategies that leverage new actions are initially unknown, and therefore attacker policies must be learned before defender adaptation can occur. To address this challenge, we develop adversarial RL (ARL) with behavior trees (ARLBT), which uses reward behavior trees (RBTs) to define progression-based reward functions and condition attacker policies on RBT execution. The proposed RBT incorporates the new attacker actions into structured attack strategies while leveraging parallel execution to generate rewards across multiple network nodes simultaneously. We evaluate ARLBT for learning attacker policies against two baseline defender policies across network scenarios with different sets of new attacker actions and compare against multiple attacker baselines. The results demonstrate that ARLBT improves attacker success compared to ARL using the original CC2 reward function and that conditioning policies on RBT execution can improve sample efficiency. We further show that defender policies can be adapted against the learned attackers, improving robustness to attacks that leverage new attacker actions.

\keywords{cyber-security \and neurosymbolic AI \and reinforcement learning \and behavior trees}
\end{abstract}

\section{Introduction}

Autonomous cyber-defense agents for computer networks trained using reinforcement learning (RL) reduce the manual effort of cyber operators by continuously monitoring networks and responding to attacks in real time. Recent work has demonstrated that RL agents can learn effective defender policies for complex enterprise-scale network environments~\cite{king2025automated, molinamarkham2025oel}. However, cyber environments are inherently non-stationary. Changes in network topology, attacker strategies, or attacker capabilities can introduce a mismatch between the conditions under which a defender was trained and the threats it must defend against during deployment~\cite{mcfadden2026sokpitfallsdeepreinforcement}. As attacker capabilities evolve and enable new strategies for disruption, evasion, and lateral movement, the effectiveness of defender policies trained against fixed attacker capabilities may degrade.

Prior work has explored adapting defender policies to changing attacker behavior and network structure. Existing approaches have considered switching between defender policies in response to changing fixed attacker policies~\cite{potteiger2024ebts, samaddar2024ood}, generalization across varying network topologies~\cite{king2025automated, molinamarkham2025oel}, and jointly training attackers and defenders through adversarial RL (ARL), where attacker and defender policies are trained against one another to improve their respective policies~\cite{multiagentbattlesim2022, coadaptation2026}. However, these particular methods assume that attacker capabilities remain fixed. In practice, attacker capabilities may evolve and enable new attack strategies that were not represented during defender training. Existing methods therefore do not address how to learn attackers capable of leveraging new capabilities or how to subsequently adapt defender policies against the resulting attacker behaviors.

In this work, we consider the problem of adapting defender policies from an original network scenario to an extended network scenario with new attacker capabilities represented by new attacker actions. Specifically, we study CAGE Challenge Scenario 2~\cite{kiely2023autonomous} (CC2), where a defender policy is initially trained against fixed attackers operating under the original set of attacker actions in an enterprise network environment. The objective of the attacker is to compromise and persistently impact an operational server, while the objective of the defender is to mitigate the attacker's progress and minimize quality of service disruption. We extend the scenario with new attacker actions that enable additional disruption and evasion capabilities, such as denial-of-service attacks and stealthy exploitation. Consequently, effective attacker policies that leverage the new actions must first be learned before defender policy adaptation can occur.

Learning attacker policies for new attacker actions presents several challenges. As the set of attacker actions expands, effective attack strategies may require progression through multiple coordinated stages over long time horizons. For example, denial-of-service may be used to create diversions before progressing toward the objective of the attacker. These long-horizon dependencies produce delayed and sparse reward signals that reduce sample efficiency, highlighting the importance of reward shaping for attacker training~\cite{rewardshapingcyborg2023, icarte2022rewardmachine, rewardmachinescyber2024}. However, the reward function for CC2 provides limited feedback regarding attacker progression. Prior work has explored finite-state machines as progression-based reward functions (reward machines) and conditioned attacker policies on attacker progression~\cite{rewardmachinescyber2024}, while more recent work has introduced reward behavior trees as progression-based reward functions for compositional robotic tasks~\cite{potteiger2026reward}.

To address these challenges, we develop adversarial reinforcement learning with behavior trees (ARLBT) to learn attacker policies for new attacker actions and subsequently adapt defender policies. We use reward behavior trees (RBTs) to define a progression-based reward function and condition attacker policies on attack progression. The RBT integrates new attacker actions into structured attack progression, while parallel execution enables simultaneous reward generation across multiple network nodes. The learned attacker policies are trained against defender policies and are subsequently used to adapt defender policies in the extended network scenario.

The technical contributions of this work are:
\begin{enumerate}
\item A method for conditioning RL policies on RBT execution by conditioning policies on disjoint sets of RBT behaviors.

\item A reward behavior tree (RBT) for attacker training in the extended network scenario that incorporates new attacker actions and uses parallel execution.

\item Adversarial reinforcement learning with behavior trees (ARLBT), an adversarial RL algorithm that uses RBT-conditioned policies and the RBT to learn attacker policies for new attacker actions and subsequently adapt defender policies in the extended network scenario.

\item An evaluation with code\footnote{\href{https://github.com/npotteig/Robust_CC2}{https://github.com/npotteig/Robust\_CC2}} across network scenarios with different sets of new attacker actions, where ARLBT attacker policies are trained and evaluated against two baseline defender policies and compared against multiple attacker baselines. The results demonstrate that ARLBT improves attacker success compared to ARL using the original CC2 reward function and that conditioning policies on the RBT can improve sample efficiency. We additionally show that defender policies can be adapted against the learned attackers, significantly improving defender reward and reducing attacker success while maintaining robustness against the original fixed attackers.

\end{enumerate}

\section{Related Work}
\noindent
RL has increasingly been applied to autonomous computer network defense. However, deploying RL agents in network defense introduces several significant challenges. A recent survey by McFadden \emph{et al.} (2026) highlights pitfalls including mismatches between training and deployment environments and non-stationarity caused by changes in attacker strategies or network topologies~\cite{mcfadden2026sokpitfallsdeepreinforcement}.

Prior work has explored adapting defender agents to changing attacker policies and network conditions. Potteiger \emph{et al.} (2024) and Samaddar \emph{et al.} (2025) use behavior trees with learned behaviors to dynamically switch between defense policies in response to changes in attacker behavior in CC2~\cite{potteiger2024ebts, samaddar2024ood}. Wolk \emph{et al.} (2022) further demonstrate that even minor modifications to the network, such as changes in node ordering or edge connectivity, can significantly degrade defender policy performance~\cite{wolk2022beyond}. To improve generalization, Molina-Markham \emph{et al.} (2025) use curriculum learning in FARLAND~\cite{MolinaMarkham2021NetworkED}, an RL simulation and emulation environment for training defenders, to train defender policies across increasingly difficult fixed attacker policies and network scales~\cite{molinamarkham2025oel}. 
King \emph{et al.} (2026) use an inductive graph neural network approach to generalize to larger network topologies while maintaining defender policy performance~\cite{king2025automated}.
Although these works consider evolving network topology, they do not address adapting to new attacker actions or learning attacker policies.

Prior work has explored adversarial RL for learning attacker and defender policies jointly. Hicks \emph{et al.} (2026) argue that training against fixed attacker policies can produce brittle defender policies and suggest co-evolving the attacker and defender to improve robustness~\cite{hicks2026buildingbetter}. Kunz \emph{et al.} (2022) extend CyberBattleSim~\cite{cyberbattlesim}, an RL training environment for attacker training, to support joint defender and attacker policy learning~\cite{multiagentbattlesim2022}. They demonstrate that learned defender policies can adapt and significantly reduce attacker progress. Similarly, Makarem \emph{et al.} (2026) jointly train defender and attacker agents in an attack graph environment using Q-learning and DQN~\cite{coadaptation2026}. In these works, the set of available attacker actions remains fixed, which may limit applicability when attackers evolve their strategies when new actions are introduced.

Reward shaping remains a major challenge in RL for cybersecurity to efficiently train defenders and attackers. Bates \emph{et al.} (2023) show that reward magnitude, sign, and curiosity-driven rewards in CC2 can affect sample efficiency~\cite{rewardshapingcyborg2023}. For attacker training, reward machines have been used to model progress through subtasks using finite-state machines as reward logic~\cite{icarte2022rewardmachine}. Li \emph{et al.} (2024) apply reward machines to guide attacker policies in CyberBattleSim and condition policies on reward machine state to improve sample efficiency~\cite{rewardmachinescyber2024}. However, reward machines lack the modularity and parallel execution naturally supported by behavior trees~\cite{potteiger2026reward, iovino2025comparingbtsandfsms}. Recent work by Potteiger \emph{et al.} (2026) introduces BT-based reward functions for compositional robotic tasks~\cite{potteiger2026reward}. Their BTs are generated automatically using an LLM with a BT template for reward shaping and demonstrate effective guidance of robotic agents under adversarial conditions. However, their formulation does not leverage BT parallel execution or condition policies on BT execution.

\section{Adaptive Cyber-Defense for Evolving Attacker Capabilities}
\noindent
We formulate the problem of learning attacker policies that leverage new attacker capabilities for subsequent defender adaptation. We first describe the original network scenario (CC2) and then introduce new attacker actions that represent these capabilities in an extended network scenario. In the extended network scenario, effective attacker policies for the new actions are initially unknown and must be learned before defender policies can adapt through adversarial training.

\begin{figure*}[t]
    \centering
    \includegraphics[width=1.0\linewidth]{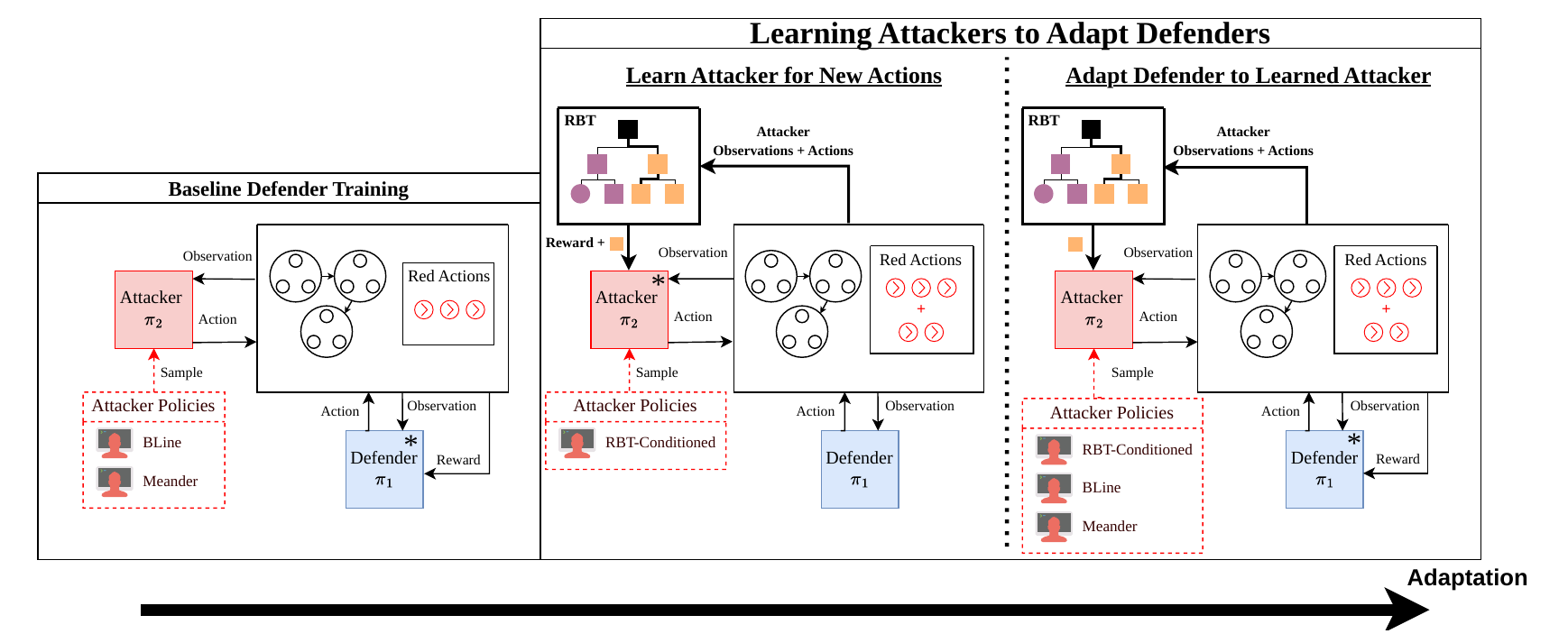}
    \caption{Overview of the proposed approach for learning attacker policies and adapting defender policies. Baseline defender policies are trained in the original network scenario and transferred to an extended network scenario with new attacker actions. Attacker policies are learned against the baseline defenders using ARLBT, where an RBT provides progression-based rewards and conditions attacker policies on attack progression. The learned attacker policies are subsequently used to adapt the defender policies in the extended network scenario. The agent being optimized at each stage is indicated by (\textbf{*}).}
    \label{fig:approach}
\end{figure*}

\subsection{Original Network Scenario: \textit{CAGE Challenge Scenario 2}}
CC2 models an enterprise network of hosts and servers with two agents: a defender and an attacker.


\subsubsection{Network Topology}
The network consists of three subnets: a user subnet with five non-critical hosts, an enterprise subnet with three servers that support services in the user subnet and a defender host, and an operational subnet with an operational server and three operational hosts. The operational server runs a critical service for the network.

\subsubsection{Agent Objectives and Actions}\label{sec:obj_actions}
The objective of the defender is to prevent the critical service from being impacted and minimize quality of service disruption. The defender can execute actions on any host or server to mitigate the attacker. The defender actions are to monitor network connections and malicious processes (\emph{Monitor}), analyze information from files associated with a host or server (\emph{Analyze}), deploy one of seven decoy services on a host or server that trigger alerts if the attacker attempts to exploit (\emph{DeployDecoy}), remove any malicious processes or files (\emph{Remove}), and restore a host to a known safe state (\emph{Restore}).


The objective of the attacker is to impact the critical service on the operational server. The attacker starts on one of the user hosts.
The attacker actions are to discover hosts and servers on a subnet (\emph{DiscoverRemoteSystems}), discover vulnerable network services on a host or server (\emph{DiscoverNetworkServices}), exploit one of the vulnerable network services (\emph{ExploitRemoteService}) establishing a low privilege session with a detection rate of $95$\% from \emph{Monitor}, and then execute privilege escalation to establish a high privilege session (\emph{PrivilegeEscalate}). 
If privilege escalation is executed on the operational server successfully, the attacker can impact the critical service (\emph{Impact}) achieving its objective.

\subsubsection{Known Attacker Policies}
Defender policies for CC2 are designed to mitigate two fixed attacker policies:  \emph{BLine} and \emph{Meander}. The \emph{BLine} policy follows a greedy path by selecting actions that most directly leads to \emph{Impact} the operational server. The \emph{Meander} policy  exploits and escalates privileges across all hosts and servers within a subnet before progressing to the next, ultimately targeting the operational server.

\subsection{Extended Network Scenario with New Attacker Actions}

\subsubsection{New Attacker Actions}
We introduce two additional attacker actions: \emph{DenialofService (DoS)}, which persistently floods network services on a host or server, and \emph{StealthExploitRemoteService}, which has a lower detection rate of $10$\%. These actions are used to disrupt services, misdirect the defender, and evade detection.

\subsubsection{Discovering Attacker Policies for New Actions}
Unlike CC2, which contains known attacker policies for designing defender policies, the extended network scenario introduces attacker actions for which effective strategies are unknown. 
These new actions enable new attack strategies that existing defender policies are not robust to.

\subsection{Modeling Network Scenarios in CybORG}

Network scenarios are simulated using CybORG, a simulation environment designed for training defender and attacker policies using RL~\cite{cyborg_acd_2021, cyborg2022}.

\subsubsection{Environment Formulation}
We model the environment as a partially observable Markov game (POMG) with two agents: a defender and an attacker. The POMG with $n=2$ agents is denoted by the tuple $\mathcal{M} = \big(\mathcal{S}$, $\{\mathcal{A}_i\}^n_{i=1}$, $\{\mathcal{O}_i\}^n_{i=1}$, $\mathbb{T}$, $\mathbb{O}$, $\{r_i\}^n_{i=1}\big)$. $\mathcal{S}$ is the state space, $\mathcal{A}_i$ is the action space for agent $i$, and $\mathcal{O}_i$ is the observation space for agent $i$. We denote the joint action as $\textbf{a} := (a_1, a_2)$ for the defender and attacker. $\mathbb{T}(s^{t+1} \mid s^t, \mathbf{a}^t)$ is the transition function to the next state $s^{t+1}$ given a joint action $\mathbf{a}^t$ is taken from the current state $s^t$ at timestep $t$. We denote the joint observation as $\mathbf{o} := (o_1, o_2)$ for the defender and attacker. $\mathbb{O}(\mathbf{o} \mid s)$ converts a state $s$ to a joint observation $\mathbf{o}$.
The state-action history up until timestep $t$ is defined as \(h^t_s = (s^1, \mathbf{a}^1, \dots, s^t)\)
and the observation-action history for agent $i$ is defined as \(h^t_i = (o^1_i, a^1_i, \dots, o^t_i)\).
Finally, $r_i$ denotes the reward function for agent $i$. In the fully observable setting, the reward is defined as $r_i(h_s^t)$. In the partially observable setting, the reward is defined as $r_i(h_i^t)$. Because these reward functions depend on histories rather than only the current state and action, they are non-Markovian. This formulation enables the reward function to capture temporal dependencies that cannot be represented using current information alone.

\subsubsection{State and Observation Spaces}
The state of the network scenario is modeled as an attributed graph with a set of nodes $\mathcal{V}$ and a set of edges $\mathcal{E} = \{ (u, v) \mid u, v \in \mathcal{V} \}$. Nodes represent hosts, servers, subnet routers, processes, and files while edges denote ownership. Each node in the attributed graph has features, $f: \mathcal{V} \rightarrow \mathbb{R}^d$, where all node features are denoted as $\mathbf{X}$. Node features represent system information (i.e. operating system, user or server, etc.), file and process information (i.e. name, type, signature, etc.), and network connections from external processes. 

The defender and attacker observation spaces, $\mathcal{O}_1$ and $\mathcal{O}_2$, provide partial views of the true state $s \in \mathcal{S}$. An observation for agent $i$ at timestep $t$ is defined as $o_i^t = (\mathcal{V}_i^t, \mathcal{E}_i^t, \mathbf{X}_i^t)$, where $\mathcal{V}_i^t \subseteq \mathcal{V}$, $\mathcal{E}_i^t \subseteq \mathcal{E}$, and $\mathbf{X}_i^t$ denotes the observable node features. The defender observation $o_1^t \in \mathcal{O}_1$ contains system information for all hosts, servers, and subnet nodes present on the network, as well as processes, files, and network connections discovered through the \emph{Monitor} and \emph{Analyze} actions. The initial attacker observation $o_2^0 \in \mathcal{O}_2$ contains the system information, files, and processes associated with the host node on which the attacker is initialized. The attacker must then expand its observable sets $\mathcal{V}_2^t$, $\mathcal{E}_2^t$, and $\mathbf{X}_2^t$ through its actions. For example, the \emph{DiscoverRemoteSystems} action adds newly discovered host and server nodes to the attacker's observation.

\subsubsection{Action Spaces} The defender and attacker can take network wide actions or actions targeted at a host or server (refer to Section~\ref{sec:obj_actions}). The defender action space $\mathcal{A}_1$ consists of an additional \emph{Sleep} action. The \emph{Monitor} action is executed in parallel with each selected defender action. The attacker action space $\mathcal{A}_2$ consists of \emph{DiscoverRemoteSystems} and two additional actions: \emph{Sleep} and \emph{ProgressAttack}. \emph{ProgressAttack} advances an attack on a host or server according to the current compromise level:
\[
\begin{aligned}
\emph{DiscoverNetworkServices}
&\rightarrow
\emph{ExploitRemoteService} \\
&\rightarrow
\emph{PrivilegeEscalate}
\rightarrow
\emph{Impact}.
\end{aligned}
\]

\subsubsection{Reward Functions}\label{sec:cc2_rew}
The fully observable defender reward function, $r_1(h^t_s)$, outputs penalties when the attacker exploits and maintains access on hosts and servers, as well as when the attacker successfully executes \emph{Impact} on the operational server. Additionally, the defender receives a penalty for executing \emph{Restore}, since this action disrupts the quality of service for benign users. The partially observable attacker reward function, $r_2(h^t_2)$, outputs the defender penalties, except for the penalty associated with the \emph{Restore} action. 


\subsection{Extending the Attacker Action Space for New Actions}
As new attacker actions are introduced, the original POMG $\mathcal{M}^0$ shifts to a POMG $\mathcal{M}^1$ with an extended attacker action space $\mathcal{A}_2^1$, where $\mathcal{A}_2^1 = \mathcal{A}_2^0 \cup \mathcal{A}_2^+$. $\mathcal{A}_2^+$ denotes a set of new actions.

$\mathcal{A}_2^+$ contains two new actions: \emph{EnableDoS} and \emph{DisableDoS}. The \emph{EnableDoS} (\emph{DisableDoS}) action enables (disables) the \emph{DoS} action on a host or server. While \emph{DoS} is enabled on a host or server, the \emph{DiscoverRemoteSystems} and \emph{ProgressAttack} actions are disabled for that host or server to model network bandwidth saturation. In \emph{ProgressAttack}, \emph{ExploitRemoteService} is upgraded to use \emph{StealthExploitRemoteService} to lower detection.

\subsection{Learning Attacker Policies to Adapt Defender Policies}

We consider a \emph{Baseline} defender policy $\pi_1(\cdot \mid o_1, \theta_1)$ with parameters $\theta_1$ that maximizes defender reward against \emph{BLine} and \emph{Meander} attacker policies in $\mathcal{M}^0$. The objective is to transfer $\pi_1$ to $\mathcal{M}^1$ such that the transferred defender policy remains robust against an optimal attacker policy $\pi_2^*$. However, since $\pi_2^*$ is not known \emph{a priori}, it must first be learned. 

Parameters $\theta_2$ for an attacker policy $\pi_2(\cdot \mid o_2, \theta_2)$ can be optimized via adversarial RL. However, learning effective attacker policies presents several challenges. The attacker reward function in CybORG primarily models compromise on hosts or servers and impact on the operational server. As the attacker action space expands, effective attack strategies may require progression through multiple stages over long time horizons. 
These long-horizon dependencies produce delayed reward signals, reducing sample efficiency and increasing the difficulty of discovering effective attacker policies.

Additionally, the original attacker reward function provides limited interpretability regarding attacker progression during execution. In network scenarios, attacker behavior naturally progresses through stages such as reconnaissance, privilege escalation, lateral movement, and impact. Without progression-based reward signals aligned with these stages, learned attacker policies lack explicit guidance regarding advancement towards the attacker objective, making it more difficult to efficiently discover effective long-horizon attack strategies.

Therefore, learning attacker policies for new attacker actions requires progression-based reward functions that both improve learning efficiency and provide interpretable representations of attacker progression towards its objective.

\begin{figure*}[t]
\centering

\begin{minipage}[c]{1.0\textwidth}
    \centering
    \includegraphics[width=\linewidth]{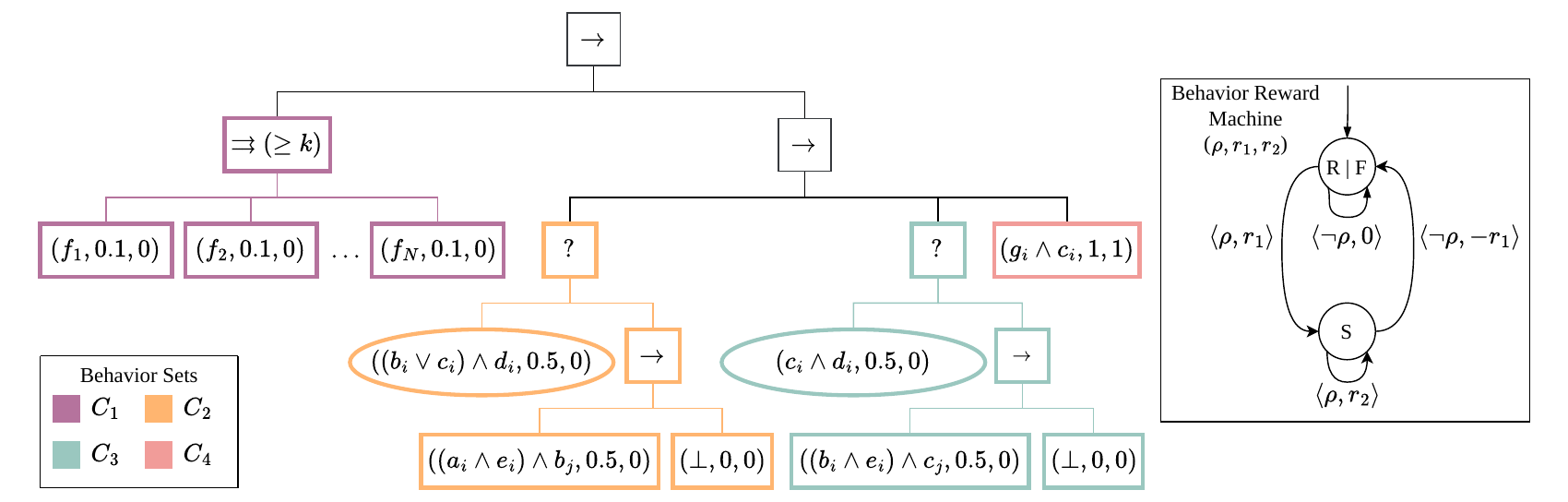}
\end{minipage}

\par\vspace{0.75em}

\begin{minipage}[c]{0.5\textwidth}
    \centering
    
    \scriptsize
    \begin{tabular}{m{0.5\linewidth} m{0.55\linewidth}}
    \hline
    \textbf{Node Proposition ($\rho$)} & \textbf{Description} \\ \hline
    $a$ & is user host \\
    $b$ & is enterprise server \\
    $c$ & is operational server \\
    $d$ & is exploited \\
    $e$ & is escalated privilege \\
    $f$ & is under DoS \\
    $g$ & \emph{Impact} executed \\ \hline
    \end{tabular}
\end{minipage}
\caption{The figure shows the attacker RBT with behavior sets $C_1, C_2, C_3$ and $C_4$, and logic expressions $\rho_i$ conditioned on node $v_i \in \mathcal{V}_2$. $\rho_i$ in the figure implies $\exists v_i \in \mathcal{V}_2 \mid \rho_i$ is true. The table shows node propositions for the attacker RBT.}
\label{fig:red_rbt}
\end{figure*}

\section{Learning Attacker Policies using Adversarial RL with Behavior Trees}
\noindent
We address the problem of learning attacker policies for new attacker actions in $\mathcal{M}^1$ using adversarial RL with behavior trees (BTs). Our approach uses a reward behavior tree (RBT) to define progression-based reward functions and condition attacker policies on disjoint sets of RBT behaviors corresponding to different stages of attack progression. The modularity of BTs enables new attacker actions to be incorporated into structured stages of attack progression, while parallel execution enables progression-based rewards to be generated across multiple network nodes simultaneously. The learned attacker policies are subsequently used to adapt baseline defender policies in the extended network scenario.

\subsection{Behavior Trees}
Behavior trees (BTs) are hierarchical symbolic structures composed of behaviors organized in a tree~\cite{colledanchise2018behavior}. BTs excel at decomposing complex objectives into a sequence of subtasks to monitor progression. Behaviors can be executed, returning one of three statuses: \textit{Running}, \textit{Success}, or \textit{Failure}. These statuses determine how execution propagates through the tree and which behaviors are executed next. There are two types of behaviors: internal control behaviors and leaf execution behaviors. Control behaviors (e.g. sequence ($\rightarrow$), fallback ($?$), and parallel ($\rightrightarrows$)) coordinate their child behaviors by deciding which to execute next and returning a status based on child return statuses. Leaf behaviors are divided into condition ($\bigcirc$) and action ($\square$) behaviors, which semantically represent conditions being checked and actions being performed, respectively.

A BT $\tau$ with behaviors $\mathcal{B}$ has internal control behaviors $\mathcal{B}_\text{internal}$ and leaf execution behaviors $\mathcal{B}_\text{leaf}$. At timestep $t$, the BT control flow executes a sequence of leaf behaviors $\varepsilon^t = (b_1, \dots, b_e)$ updating their statuses and backpropagating statuses to internal behaviors.

\subsection{Behavior Trees as Reward Functions}

Reward behavior trees (RBTs) were introduced in prior work~\cite{potteiger2026reward} as a method for defining progression-based reward functions using behavior trees (BTs). RBTs represent leaf behaviors using reward machines, allowing agent progression to be represented as transitions between behaviors while simultaneously producing reward signals aligned with progression through the BT. In this work, we extend the use of RBTs by conditioning RL policies on RBT execution and develop an RBT for attacker training that uses parallel execution to generate progression-based rewards across multiple network nodes.

A reward machine (RM) is a finite-state machine that inputs abstract environment descriptions and outputs reward functions~\cite{icarte2022rewardmachine}. A simple RM restricts this output to a constant scalar reward. Given a set of propositions $\mathcal{P}$, a simple RM is defined by the tuple $\langle U, u^0, \delta_u, \delta_r \rangle$, where $U$ is a finite set of machine states, $u^0 \in U$ is the initial state, $\delta_u: U \times 2^\mathcal{P} \rightarrow U$ is the state-transition function, and $\delta_r: U \times 2^\mathcal{P} \rightarrow \mathbb{R}$ is the state-reward function.

Propositions are evaluated using a labeling function $\ell_i: \mathcal{O}_i \times \mathcal{A}_i \rightarrow 2^\mathcal{P}$, which inputs observation $o^t_i$ and previous action $a^{t-1}_i$ at timestep $t$ and outputs a set of true propositions $\sigma^t \in 2^\mathcal{P}$.  
Given a label history at timestep $t$, $\lambda^t = \langle \sigma^0, \dots, \sigma^t \rangle \in (2^\mathcal{P})^+$, the RM transitions to state $u^{t} = \delta_u(u^{t-1}, \sigma^t)$ and outputs a reward using
\(
\delta_r(u^{t-1}, \sigma^t).
\)

To align RM execution with BT semantics, the RMs in an RBT share the set of states
\(
U = \{\texttt{Running}, \texttt{Failure}, \texttt{Success}\},
\)
which correspond directly to BT return statuses. The initial states for condition and action behaviors are $\texttt{Failure}$ and $\texttt{Running}$, respectively. This correspondence allows RMs to function directly as BT leaf behaviors, where the current RM state is returned as the behavior status.

Given a shared proposition set $\mathcal{P}$ and labeling function $\ell$, the RBT $\tau_R$ maintains the current RM state for each leaf behavior $b \in \mathcal{B}_{\text{leaf}}$. At timestep $t+1$, the executed leaf RMs $\varepsilon^{t+1}$ transition state according to
\begin{equation}
u^{t+1}(b) =
\begin{cases}
\delta_u^b\big(u^t(b), \sigma^{t+1}\big) & \text{if } b \in \varepsilon^{t+1} \\
u^t(b) & \text{o/w.}
\end{cases}
\end{equation}
The RBT reward is defined as the sum of rewards produced by the executed leaf RMs:
\begin{equation}
R_{\tau_R}(\lambda^{t+1}) = \sum_{b \in \varepsilon^{t+1}} \delta_r^b(u^t(b), \sigma^{t+1}).
\end{equation}
The RBT reward depends on the label history $\lambda^t$ derived from the observation-action history $h_i^t$ for agent $i$. Therefore, the RBT acts as a partially observable reward function $r_i(h^t_i)$. 

\subsection{Conditioning RL Policies on Reward Behavior Trees}

The execution progress of an RBT $\tau_R$ can be leveraged to condition a policy $\pi_i(\cdot \mid o_i, \theta_i)$ for agent $i$. Let
$
\mathcal{C} = \{C_1, \dots, C_m\} \subseteq 2^\mathcal{B}
$
be a collection of $m$ disjoint behavior sets.
Each behavior set $C_j$ may correspond to a single leaf behavior or a subtree within the RBT. The RBT-conditioned policy
$
\pi_i(\cdot \mid o_i, C_j, \theta_i),
$
for a behavior set $C_j \in \mathcal{C}$
provides progression-based context regarding the current execution progress of the RBT. To select a behavior set, let
$
\phi_{\tau_R} : (2^\mathcal{P})^+ \rightarrow \mathcal{C}
$
be a selection function that maps a label history $\lambda \in (2^\mathcal{P})^+$ to a behavior set $C_j \in \mathcal{C}$. At timestep $t$, the RBT executes a sequence of leaf behaviors, $\varepsilon^t$, and the selected behavior set is determined according to the final executed leaf behavior $b_e \in \varepsilon^t$ such that
$
\phi_{\tau_R}(\lambda^t) = C^t = C_j \in \mathcal{C}
\quad \text{where} \quad
b_e \in C_j.
$

\subsection{Designing an Attacker Reward Behavior Tree}

We design an RBT $\tau^2_R$ that considers progression to impact the operational server using the new attacker actions in $\mathcal{M}^1$. The RBT is divided into two primary stages for the attacker to flood the network using the \emph{DoS} action, then propagate through the network to impact the operational server using the \emph{StealthExploitRemoteService} action. The RBT is parameterized by an upper bound $N$ on the number of nodes in the network and a threshold $k$ denoting the minimum number of nodes on which \emph{DoS} must be enabled. 
The set of propositions $\mathcal{P}$ contain node-wise propositions in Figure~\ref{fig:red_rbt} for each known node $v_i \in \mathcal{V}_2$. The size of $\mathcal{P}$ increases proportionally to $\mathcal{V}_2$ as more nodes are discovered.

The RBT illustrated in Figure~\ref{fig:red_rbt}, consists of $m=4$ behavior sets $\mathcal{C} = \{C_1, C_2, C_3, C_4\}$. $C_1$ is a subtree under the root (sequence) that tracks if \emph{DoS} is enabled on at least $k$ nodes. This subtree encourages an attacker to select actions to flood the network, thereby distracting the defender. It is implemented as a parallel behavior with $N$ children, each monitors a node and outputs a reward when \emph{EnableDoS} is executed on that node. The parallel behavior returns \texttt{Running} until at least $k$ children return \texttt{Success}, at which point it returns \texttt{Success}. The second subtree under the root (sequence) contains $C_2, C_3, C_4$, and tracks progress towards the objective of the attacker. $C_2$ tracks propagation from the user subnet to the enterprise subnet using a fallback-sequence structure with three leaf behaviors (one condition and two actions). Specifically, the agent must first discover a server in the enterprise subnet while maintaining a privileged foothold in the user subnet, and then execute \emph{StealthExploitRemoteService} to exploit an enterprise server. The second action executes until the enterprise server is exploited. $C_3$ mirrors this structure, but instead tracks propagation from the enterprise subnet to the operational server. $C_4$ tracks persistent execution of the \emph{Impact} action on the operational server. Each leaf behavior is associated with an RM conditioned on a proposition $\rho$, with a primary reward $r_1$ for transitions between different RM states and a reward $r_2$ for remaining in the \texttt{Success} state. 

\subsection{Adversarial RL with Behavior Trees}

We design an adversarial RL algorithm that uses the RBT (Figure~\ref{fig:red_rbt}) to optimize parameters $\theta_2$ for a RBT-conditioned attacker policy $\pi_2(\cdot \mid o_2, C_j, \theta_2)$ in $\mathcal{M}^1$ against a pretrained defender policy $\pi_1(\cdot \mid o_1, \theta_1)$. 

Algorithm~\ref{alg:rbtarl} summarizes adversarial RL with behavior trees (ARLBT). For each training iteration, trajectories are collected in a POMG environment $\mathcal{M}$ using the current defender and attacker policies. At each timestep, the behavior-set selection function $\phi_{\tau_R}$ maps the current label history to a behavior set used to condition the attacker policy. The defender and attacker actions are jointly executed in the $\mathcal{M}$, after which the state-action and label histories are updated and rewards are computed. The defender reward is obtained from the fully observable reward function $r_1(h_s^t)$, while the attacker reward is computed using the RBT reward function $R_{\tau_R}(\lambda_t)$. The resulting transitions are stored in replay buffers for both agents. After trajectory collection, proximal policy optimization (PPO)~\cite{schulman2017proximalpolicyoptimizationalgorithms} is used to update agent $i$.

The attacker parameters $\theta_2$ are first optimized in $\mathcal{M}^1$ by setting the agent $i$ to \emph{attacker}. Subsequently, the \emph{Baseline} defender parameters $\theta_1$ can be optimized against the learned attacker policy in $\mathcal{M}^1$ by setting the agent $i$ to \emph{defender}. During defender training, the \emph{BLine} and \emph{Meander} attacker policies can additionally be sampled alongside the RBT-conditioned attacker policy to maintain robustness against previously observed attacker behaviors.

\begin{algorithm}[t]
\small
\caption{Adversarial Reinforcement Learning with Behavior Trees (ARLBT)}
\label{alg:rbtarl}
\KwIn{Defender parameters $\theta_1$, attacker parameters $\theta_2$, POMG $\mathcal{M}$, attacker RBT $\tau_R$, labeling function $\ell_2$, agent $i$, training iterations $I$, episodes $E$, timesteps $T$}
\KwOut{Updated $\theta_1, \theta_2$}

\For{$j= 1, \dots, I$}{
Initialize replay buffers $\mathcal{D}_1 \leftarrow \{\}, \mathcal{D}_2 \leftarrow \{\}$\;

\For{$e = 1, \dots, E$}{
    Reset environment $\mathcal{M}$\;
    
    Observe $s^0, o_1^0, o_2^0$\;
    Initialize histories $h_s^0 \leftarrow (s^0), \lambda^0 \leftarrow ()$\;
    
    \For{$t = 0, \dots, T-1$}{
        $C^t \leftarrow \phi_{\tau_R}(\lambda^t)$\;
        $a_1^t \sim \pi_1(\cdot \mid o_1^t,  \theta_1)$,
        $a_2^t \sim \pi_2(\cdot \mid o_2^t, C^t, \theta_2)$\;
        
        Execute $\mathbf{a}^t = (a_1^t, a_2^t)$ in $\mathcal{M}$\;
        Observe $s^{t+1}, o_1^{t+1}, o_2^{t+1}$\;
        Updates histories $h_s^{t+1} \leftarrow (h^t_s, \mathbf{a}^t, s^{t+1})$,
        $\lambda^{t+1} \leftarrow (\lambda^t, \ell_2(o_2^{t+1}, a_2^t))$\;

        Calculate Rewards 
        $r_1^{t+1} \leftarrow r_1(h_s^{t+1})$,
        $r_2^{t+1} \leftarrow R_{\tau_R}(\lambda^{t+1})$\;

        Store transitions in $\mathcal{D}_1, \mathcal{D}_2$\;
        
    }
    
}
\If{\text{agent} $i == $ defender}{
        Update $\theta_1$ using PPO and $\mathcal{D}_1$\;
}
\ElseIf{\text{agent} $i == $ attacker}{
    Update $\theta_2$ using PPO and $\mathcal{D}_2$\;
}
}

\Return $\theta_1, \theta_2$\;
\end{algorithm}

\section{Evaluation}
\noindent
We evaluate ARLBT to learn attacker policies for new attacker actions and adapting defender policies. We compare attacker policies learned with ARLBT against two defender policies across four network scenarios containing different sets of new attacker actions against three attacker baselines. We additionally evaluate whether defender policies can be adapted against attackers learned with ARLBT. All experiments are conducted on an Ubuntu 22.04 workstation equipped with an AMD Ryzen Threadripper PRO 5995WX 64-core processor with 128 hardware threads and 500 GB of memory.

\subsection{Experimental Setup}
\subsubsection{Network Scenarios}
The four ablations of new attacker actions in network scenarios are: \emph{Orig} the original network scenario (CC2), \emph{$+$DoS} an extended network scenario with \emph{DoS} as the added action, \emph{$+$Stealth} an extended network scenario with \emph{StealthExploitRemoteService} as the added action, and \emph{$+$DoS,Stealth} adds both actions.

\subsubsection{Defender Policies}
We consider two high performing \emph{Baseline} defenders in CC2. \emph{CardiffUni} is an agent that trains separate policies on \emph{BLine} and \emph{Meander} and uses a heuristic to select between them at test time~\cite{hannay2022cage2}. They trained a policy using this approach that was the winner of the competition for CC2. \emph{KEEP} is a state-of-the-art agent from~\cite{king2025automated} that uses a graph neural network with global attention layers to generalize across varying network topologies.

\subsubsection{Attacker Policies}
\emph{BLine} and \emph{Meander} are the fixed policies in CC2. \emph{ARL (CC2)} is a policy learned via adversarial RL using the CC2 attacker reward function in Section~\ref{sec:cc2_rew}. \emph{ARLBT (Reward Only)} is a policy learned via an ablation of ARLBT without conditioning on the RBT. \emph{ARLBT (Full)} is an RBT-conditioned policy learned via ARLBT.

\subsubsection{Success Metrics}
The success metrics include \emph{attacker success} and cumulative \emph{defender reward} per episode. \emph{Attacker success} is the proportion of timesteps in an episode during which \emph{Impact} was successfully executed on the operational server. This metric provides insight into the efficiency and frequency of the attacker achieving its objective.

\subsubsection{Model Architecture}
The attacker policy uses an actor-critic architecture with graph neural networks and global attention layers. Observations are processed using two graph convolutional network (GCN) layers. Global attention layers, originally developed for the \emph{KEEP} agent, are interleaved between the GCN layers to produce a global context vector representing the network state. This global vector is concatenated with the node embeddings produced by the GCN layers.

For the critic architecture, the global context vector and node embeddings are input to a fully connected layer that outputs a scalar value estimate. For the actor architecture, the global context vector and node embeddings are input to a fully connected layer that outputs action logits over all nodes, assuming a known upper bound on the number of nodes. Outputs corresponding to undiscovered nodes are masked in both the actor and critic architectures, restricting attacker policy decision-making to observations from known nodes in $\mathcal{V}_2$. All layers are 64-dimensional.

\subsubsection{Training Hyperparameters}

For the attacker RBT, $N=13$ and $k=3$ requiring about $23\%$ of hosts and servers to have \emph{DoS} enabled for the \emph{$+$DoS} and \emph{$+$DoS,Stealth} network scenarios. $k=0$ for \emph{Orig} and \emph{$+$Stealth}. Training iterations are $I = 500$, episodes are $E=25$, and timesteps are $T=100$. Learning rates for the attacker actor and critic are $3e-4$ and $1e-3$. 

\subsection{Attacker Success with New Actions}

\begin{table*}[t]
\caption{Attacker success of different attacker policies evaluated against
\emph{Baseline} defenders across different network scenarios, each with
$1000$ episodes. The row-wise best mean attacker success is shown in bold.}
\label{tab:red_eval_results}
\setlength{\tabcolsep}{2.5pt}
\scriptsize
\centering
\begin{tabular}{llccccc}
\toprule
Defender & Scenario & BLine & Meander & ARL (CC2) & ARLBT (Reward Only) & ARLBT (Full) \\
\midrule

\multirow{4}{*}{CardiffUni}
& \emph{Orig}
& \multirow{4}{*}{$0$}
& \multirow{4}{*}{$0$}
& $0.03$
& $0.65$
& $\mathbf{0.66}$ \\

& \emph{$+$DoS}
&
&
& $0.04$
& $0.58$
& $\mathbf{0.61}$ \\

& \emph{$+$Stealth}
&
&
& $0.07$
& $\mathbf{0.72}$
& $0.72$ \\

& \emph{$+$DoS,Stealth}
&
&
& $0.24$
& $\mathbf{0.73}$
& $0.67$ \\

\midrule

\multirow{4}{*}{KEEP}
& \emph{Orig}
& \multirow{4}{*}{$0$}
& \multirow{4}{*}{$0$}
& $0$
& $0$
& $0$ \\

& \emph{$+$DoS}
&
&
& $0$
& $0.42$
& $\mathbf{0.43}$ \\

& \emph{$+$Stealth}
&
&
& $0.14$
& $\mathbf{0.73}$
& $0.71$ \\

& \emph{$+$DoS,Stealth}
&
&
& $0.15$
& $0.60$
& $\mathbf{0.63}$ \\

\bottomrule
\end{tabular}
\end{table*}

\begin{figure}
    \centering
    \subfloat[\emph{CardiffUni}\label{fig:cardiff_red_train}]{
        \includegraphics[width=0.47\linewidth]{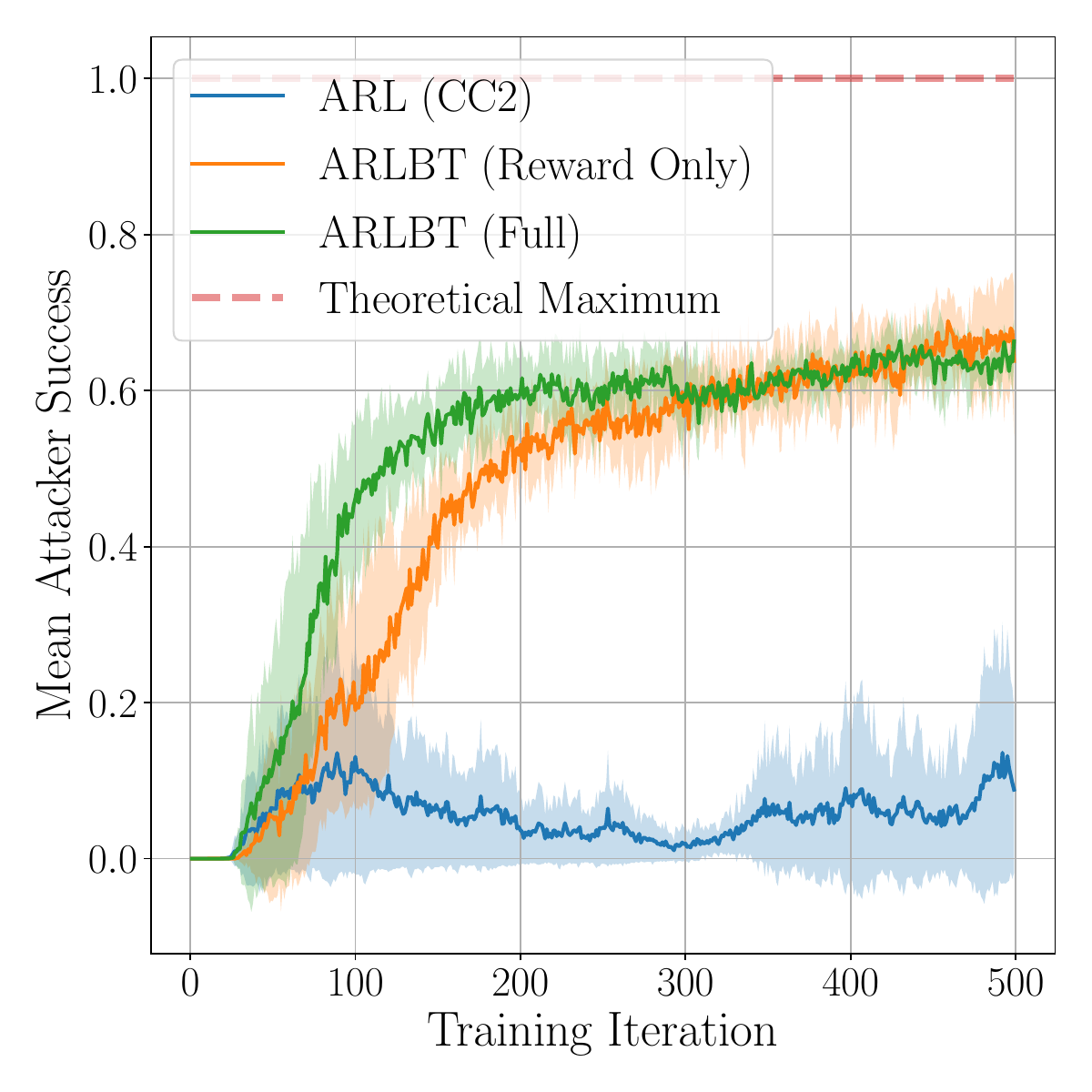}
    }
    \hfill
    \subfloat[\emph{KEEP}\label{fig:keep_red_train}]{
        \includegraphics[width=0.47\linewidth]{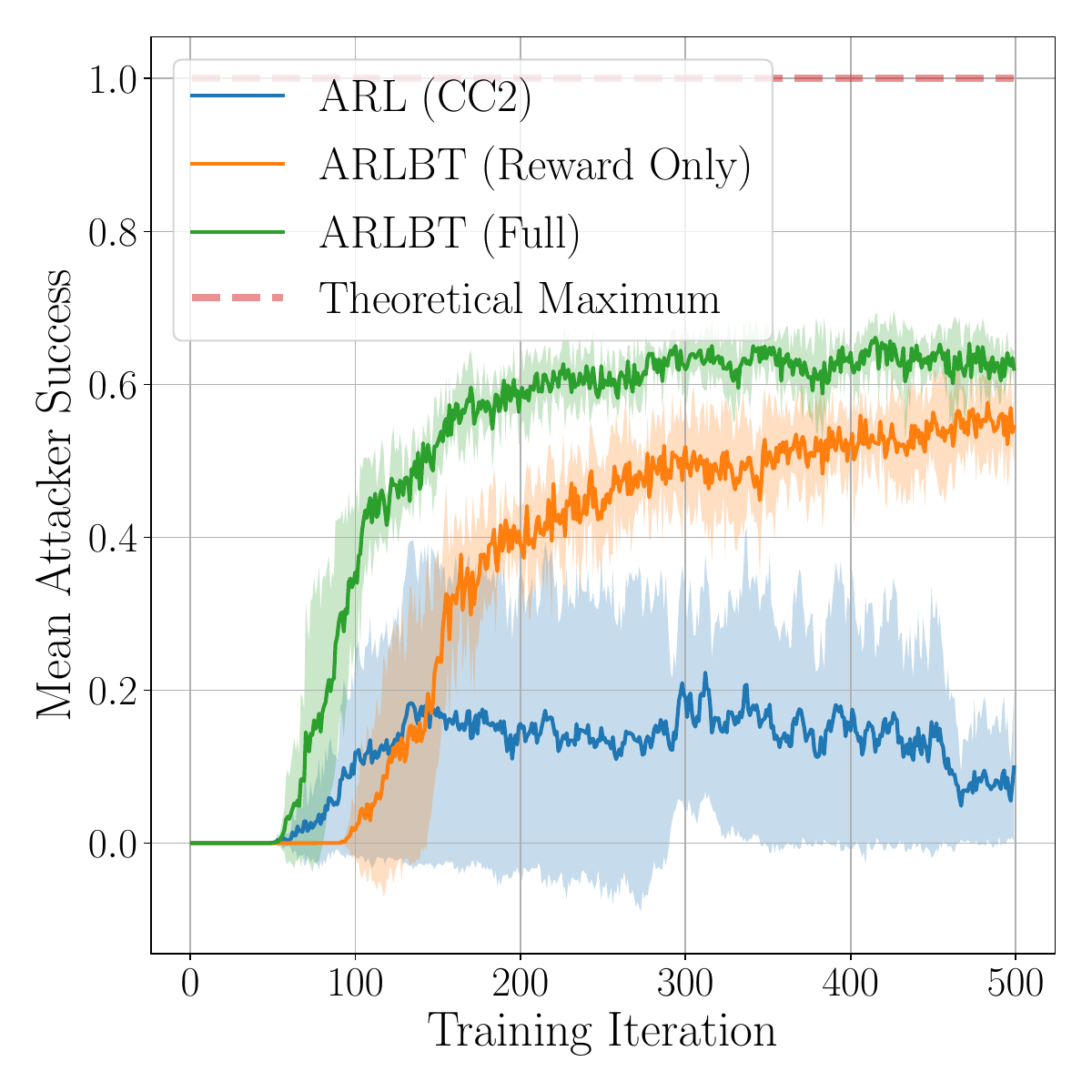}
    }
    \caption{Training curves for attacker policies evaluated against baseline defenders in the extended network scenario.}
    \label{fig:red_training_curves}
\end{figure}

We train the attacker policies for four training runs against the \emph{Baseline} defenders across different network scenarios and select the policy with the highest mean \emph{attacker success}.

\subsubsection{Testing Performance}

We evaluate whether attacker policies can maximize \emph{attacker success}. The results in Table~\ref{tab:red_eval_results} show that \emph{ARLBT (Full)} and \emph{ARLBT (Reward Only)} signifcantly increase \emph{attacker success} compared to the fixed policies, \emph{BLine} and \emph{Meander}, and the learned \emph{ARL (CC2)} policy in the network scenarios with new actions. Similarly, all learned attacker policies significantly decrease the \emph{defender reward} compared to the fixed attacker policies. Although \emph{ARL (CC2)} decreases \emph{defender reward}, its lack of explicit progression towards the attacker objective results in lower \emph{attacker success} than the ARLBT policies. Thus, progression-based reward functions, e.g., the RBT, for learning attackers are more effective to achieve the objective of the attacker. 

\subsubsection{Training Performance}
We evaluate sample efficiency based on \emph{attacker success} over training in the \emph{$+$DoS,Stealth} network scenario. Figure~\ref{fig:red_training_curves} shows that training \emph{ARLBT (Full)} improves sample efficiency on average over \emph{ARLBT (Reward Only)}. On average the final \emph{attacker success} is $14.5\%$ better against \emph{KEEP} and comparable against \emph{CardiffUni}. Thus, conditioning an attacker policy on RBT can improve sample efficiency over using the RBT as just a reward function.

\begin{table*}[t]
\centering
\caption{\emph{Defender reward} (\emph{attacker success}) of baseline and adapted \emph{KEEP} defender policies against attacker policies across different network scenarios each with $1000$ episodes.}
\label{tab:defender_adaptation}
\setlength{\tabcolsep}{4pt}
\scriptsize
\begin{tabular}{lcccccc}
\toprule
& \multicolumn{3}{c}{\textbf{Baseline}} & \multicolumn{3}{c}{\textbf{Adapted}} \\
\cmidrule(lr){2-4} \cmidrule(lr){5-7}
\textbf{Scenario} 
& ARLBT (Full) 
& BLine 
& Meander
& ARLBT (Full) 
& BLine 
& Meander \\
\midrule
\emph{Orig}              & $-42.1\ (0)$ & $-14.9 \ (0)$ & $-16.8\ (0)$ & $-35.9\ (0)$ & $-17.8\ (0)$ & $-21.1\ (0)$ \\
\emph{$+$DoS}         & $-392.4\ (0.43)$ & $-14.9\ (0)$ & $-16.8\ (0)$ & $-21.3\ (0)$ & $-19.1\ (0)$ & $-21.5\ (0)$ \\
\emph{$+$Stealth}     & $-757.5\ (0.71)$ & $-14.9\ (0)$ & $-16.8\ (0)$ & $-55.2\ (0.08)$ & $-24.9\ (0)$ & $-21.7\ (0)$ \\
\emph{$+$DoS,Stealth} & $-894.1\ (0.63)$ & $-14.9\ (0)$ & $-16.8\ (0)$ & $-39.7 \ (0)$ & $-21.2 \ (0)$ &  $-18.8\ (0)$ \\
\bottomrule
\end{tabular}
\end{table*}

\subsection{Adapting a Defender to a Learned Attacker}
We evaluate whether \emph{KEEP} can be optimized via ARLBT by sampling from \emph{ARLBT (Full)}, \emph{BLine}, and \emph{Meander} attacker policies during training across the network scenarios. \emph{CardiffUni} is excluded because its heuristic-based policy selection mechanism is incompatible with ARLBT without significant modification. The results in Table~\ref{tab:defender_adaptation} show that \emph{KEEP} can be adapted to network scenarios with new actions to substantially increase the \emph{defender reward} and reduce \emph{attacker success} against \emph{ARLBT (Full)} to near-zero. Additionally, only a marginal decrease in mean \emph{defender reward} is observed against \emph{BLine} and \emph{Meander}, suggesting that the adapted defender policy maintains robustness against the original attacker policies while adapting to the learned attacker.

\section{Conclusion}
In this work, we study adapting cyber-defense agents to evolving attacker capabilities represented by new attacker actions in an extended network scenario. We introduced adversarial reinforcement learning with behavior trees (ARLBT), which uses reward behavior trees (RBTs) to define progression-based reward functions and condition attacker policies on attack progression. Results showed that ARLBT can improve sample efficiency and learn attacker policies that achieve greater success than those trained using the CAGE Challenge Scenario 2 reward function. We further demonstrate that defender policies can be adapted against the learned attackers, improving defender reward and reducing attacker success while maintaining robustness against the original attacker policies. A limitation of ARLBT is its reliance on manually designed RBTs, which require domain expertise to define structured attack progression, although the modularity of behavior trees enables their extension and modification. Future work includes automating RBT generation for cyber-defense and extending ARLBT to continual learning settings with evolving attacker capabilities.

\section{Acknowledgment}
\noindent
This material is based on research sponsored by DARPA under agreement number HR001124C0425; The U.S. Government is authorized to reproduce and distribute reprints for Governmental purposes notwithstanding any copyright notation thereon. The views and conclusions contained herein are those of the authors and should not be interpreted as necessarily representing the official policies or endorsements, either expressed or implied, of DARPA or the U.S. Government.
\section{Disclosure of Interests}
The authors have no competing interests to declare that are relevant to the content of this article.

\bibliographystyle{splncs04}
\bibliography{ref}

@misc{hannay2022cage2,
  author       = {John Hannay},
  title        = {CardiffUni CAGE2 Submission},
  year         = {2022},
  howpublished = {\url{https://github.com/john-cardiff/-cyborg-cage-2}},
  note         = {GitHub repository}
}

@book{colledanchise2018behavior,
  title={Behavior trees in robotics and AI: An introduction},
  author={Colledanchise, Michele and {\"O}gren, Petter},
  year={2018},
  publisher={CRC Press},
  address = {Boco Raton, FL, USA}
}

@ARTICLE{iovino2025comparingbtsandfsms,
  author={Iovino, Matteo and Förster, Julian and Falco, Pietro and Jen Chung, Jen and Siegwart, Roland and Smith, Christian},
  journal={IEEE Transactions on Automation Science and Engineering}, 
  title={Comparison Between Behavior Trees and Finite State Machines}, 
  year={2025},
  volume={22},
  number={},
  pages={21098-21117},
  doi={10.1109/TASE.2025.3610090}}

@misc{schulman2017proximalpolicyoptimizationalgorithms,
      title={Proximal Policy Optimization Algorithms}, 
      author={John Schulman and Filip Wolski and Prafulla Dhariwal and Alec Radford and Oleg Klimov},
      year={2017},
      eprint={1707.06347},
      archivePrefix={arXiv},
      primaryClass={cs.LG},
      url={https://arxiv.org/abs/1707.06347}, 
}

@INPROCEEDINGS{king2025automated,
  author={King, Isaiah J. and Bowman, Benjamin and Huang, H. Howie},
  booktitle={2026 56th Annual IEEE International Conference on Dependable Systems and Networks (DSN)}, 
  title={Generalizable Graph-based Reinforcement Learning Agents for Automated Cyber Defense}, 
  year={2026},
  volume={},
  number={},
  pages={813-824},
  doi={10.1109/DSN69566.2026.00079}}

@inproceedings{potteiger2026reward,
  author    = {Nicholas Potteiger and Ankita Samaddar and Taylor T. Johnson and Xenofon Koutsoukos},
  title     = {Reward Shaping and Action Masking for Compositional Tasks using Behavior Trees and LLMs},
  booktitle = {Proceedings of the 3rd International Conference on Neuro-Symbolic Systems (NeuS)},
  year      = {2026},
  eprint    = {2605.05795},
  archivePrefix = {arXiv},
  primaryClass = {cs.LG},
  url={https://arxiv.org/abs/2605.05795}
}

@article{icarte2022rewardmachine,
author = {Toro Icarte, Rodrigo and Klassen, Toryn Q. and Valenzano, Richard and McIlraith, Sheila A.},
title = {Reward Machines: Exploiting Reward Function Structure in Reinforcement Learning},
year = {2022},
issue_date = {May 2022},
publisher = {AI Access Foundation},
address = {El Segundo, CA, USA},
volume = {73},
issn = {1076-9757},
url = {https://doi.org/10.1613/jair.1.12440},
doi = {10.1613/jair.1.12440},
journal = {J. Artif. Int. Res.},
month = may,
numpages = {36}
}

@misc{mcfadden2026sokpitfallsdeepreinforcement,
      title={SoK: The Pitfalls of Deep Reinforcement Learning for Cybersecurity}, 
      author={Shae McFadden and Myles Foley and Elizabeth Bates and Ilias Tsingenopoulos and Sanyam Vyas and Vasilios Mavroudis and Chris Hicks and Fabio Pierazzi},
      year={2026},
      eprint={2602.08690},
      archivePrefix={arXiv},
      primaryClass={cs.LG},
      url={https://arxiv.org/abs/2602.08690}, 
}

@misc{hicks2026buildingbetter,
      title={Building Better Environments for Autonomous Cyber Defence}, 
      author={Chris Hicks and Elizabeth Bates and Shae McFadden and Isaac Symes Thompson and Myles Foley and Ed Chapman and Nickolas Espinosa Dice and Ankita Samaddar and Joshua Sylvester and Himanshu Neema and Nicholas Butts and Nate Foster and Ahmad Ridley and Zoe M and Paul Jones},
      year={2026},
      eprint={2604.08805},
      archivePrefix={arXiv},
      primaryClass={cs.CR},
      url={https://arxiv.org/abs/2604.08805}, 
}

@Article{coadaptation2026,
AUTHOR = {Makarem, Mohammed A. and Razaz, Muneef A. and Saleh, Zead},
TITLE = {Co-Adaptive Attacker–Defender Learning over Attack Graphs: A Stochastic Game Approach to Dynamic Network Defense},
JOURNAL = {Future Internet},
VOLUME = {18},
YEAR = {2026},
NUMBER = {5},
ARTICLE-NUMBER = {239},
URL = {https://www.mdpi.com/1999-5903/18/5/239},
ISSN = {1999-5903},
DOI = {10.3390/fi18050239}
}

@inproceedings{cyborg2022,
author = {Foley, Myles and Hicks, Chris and Highnam, Kate and Mavroudis, Vasilios},
title = {Autonomous Network Defence using Reinforcement Learning},
year = {2022},
isbn = {9781450391405},
publisher = {Association for Computing Machinery},
address = {New York, NY, USA},
url = {https://doi.org/10.1145/3488932.3527286},
doi = {10.1145/3488932.3527286},
booktitle = {Proceedings of the 2022 ACM on Asia Conference on Computer and Communications Security},
numpages = {3},
location = {Nagasaki, Japan},
series = {ASIA CCS '22}
}

@inproceedings{wolk2022beyond,
  author    = {Melody Wolk and Andy Applebaum and Camron Dennler and Patrick Dwyer and Marina Moskowitz and Harold Nguyen and Nicole Nichols and Nicole Park and Paul Rachwalski and Frank Rau and Adrian Webster},
  title     = {Beyond CAGE: Investigating Generalization of Learned Autonomous Network Defense Policies},
  booktitle = {NeurIPS 2022 Workshop on Reinforcement Learning for Real Life},
  year      = {2022},
  eprint    = {2211.15557},
  archivePrefix = {arXiv},
  primaryClass = {cs.LG}
}

@INPROCEEDINGS{samaddar2024ood,
  author={Samaddar, Ankita and Potteiger, Nicholas and Koutsoukos, Xenofon},
  booktitle={2025 IEEE 4th International Conference on AI in Cybersecurity (ICAIC)}, 
  title={Out-of-Distribution Detection for Neurosymbolic Autonomous Cyber Agents}, 
  year={2025},
  volume={},
  number={},
  doi={10.1109/ICAIC63015.2025.10849024}}

@inproceedings{rewardshapingcyborg2023,
author = {Bates, Elizabeth and Mavroudis, Vasilios and Hicks, Chris},
title = {Reward Shaping for Happier Autonomous Cyber Security Agents},
year = {2023},
isbn = {9798400702600},
publisher = {Association for Computing Machinery},
address = {New York, NY, USA},
url = {https://doi.org/10.1145/3605764.3623916},
doi = {10.1145/3605764.3623916},
booktitle = {Proceedings of the 16th ACM Workshop on Artificial Intelligence and Security},
pages = {221–232},
numpages = {12},
location = {Copenhagen, Denmark},
series = {AISec '23}
}

@article{kiely2023autonomous,
  title={On Autonomous Agents in a Cyber Defence Environment. arXiv 2023},
  author={Kiely, Mitchell and Bowman, David and Standen, Maxwell and Moir, Christopher},
  journal={arXiv preprint arXiv:2309.07388},
  year={2023}
}

@INPROCEEDINGS{potteiger2024ebts,
  author={Potteiger, Nicholas and Samaddar, Ankita and Bergstrom, Hunter and Koutsoukos, Xenofon},
  booktitle={2024 International Conference on Assured Autonomy (ICAA)}, 
  title={Designing Robust Cyber-Defense Agents with Evolving Behavior Trees}, 
  year={2024},
  volume={},
  number={},
  pages={1-10},
  doi={10.1109/ICAA64256.2024.00011}}

@article{MolinaMarkham2021NetworkED,
  title={Network Environment Design for Autonomous Cyberdefense},
  author={Andres Molina-Markham and Cory Miniter and Becky Powell and Ahmad Ridley},
  journal={ArXiv},
  year={2021},
  volume={abs/2103.07583},
  url={https://api.semanticscholar.org/CorpusID:232232727}
}

@misc{molinamarkham2025oel,
      title={Training RL Agents for Multi-Objective Network Defense Tasks}, 
      author={Andres Molina-Markham and Luis Robaina and Sean Steinle and Akash Trivedi and Derek Tsui and Nicholas Potteiger and Lauren Brandt and Ransom Winder and Ahmad Ridley},
      year={2025},
      eprint={2505.22531},
      archivePrefix={arXiv},
      primaryClass={cs.LG},
      url={https://arxiv.org/abs/2505.22531}, 
}

@misc{cyberbattlesim,
  author       = {{Microsoft Defender Research Team}},
  title        = {CyberBattleSim},
  year         = {2021},
  publisher    = {GitHub},
  howpublished = {\url{https://github.com/microsoft/cyberbattlesim}},
  note         = {Created by Christian Seifert, Michael Betser, William Blum, James Bono, Kate Farris, Emily Goren, Justin Grana, Kristian Holsheimer, Brandon Marken, Joshua Neil, Nicole Nichols, Jugal Parikh, and Haoran Wei}
}

@INPROCEEDINGS{multiagentbattlesim2022,
  author={Kunz, Thomas and Fisher, Christian and Novara-Gsell, James La and Nguyen, Christopher and Li, Li},
  booktitle={2022 International Conference on Computational Science and Computational Intelligence (CSCI)}, 
  title={A Multiagent CyberBattleSim for RL Cyber Operation Agents}, 
  year={2022},
  volume={},
  number={},
  pages={897-903},
  doi={10.1109/CSCI58124.2022.00161}}

@INPROCEEDINGS{rewardmachinescyber2024,
  author={Li, Yuanliang and Dai, Hanzheng and Yan, Jun},
  booktitle={2024 International Joint Conference on Neural Networks (IJCNN)}, 
  title={Knowledge-Informed Auto-Penetration Testing Based on Reinforcement Learning with Reward Machine}, 
  year={2024},
  volume={},
  number={},
  doi={10.1109/IJCNN60899.2024.10650368}}

@INPROCEEDINGS{cyborg_acd_2021,
  author = {Standen, Maxwell and Lucas, Martin and Bowman, David and Richer, Toby J\. and Kim, Junae and Marriott, Damian},
  Title = {CybORG: A Gym for the Development of Autonomous Cyber Agents},
  booktitle = {IJCAI-21 1st International Workshop on Adaptive Cyber Defense.}, 
  Year = {2021}
}

\end{document}